\documentclass
[twocolumn,showpacs,preprintnumbers,amsmath,amssymb]
{revtex4}
\usepackage{graphicx}
\usepackage{dcolumn}
\usepackage{bm}
\usepackage{here}  
\usepackage{color}
\usepackage{amsmath}
\usepackage{amssymb}
\usepackage{fancybox}

\begin{document}

\title{Scaling Properties of Dynamical Localization in Monochromatically 
Perturbed Quantum Maps: standard map and Anderson map}
\author{Hiroaki S. Yamada}
\affiliation{Yamada Physics Research Laboratory,
Aoyama 5-7-14-205, Niigata 950-2002, Japan}
\author{Fumihiro Matsui} 
\affiliation{Department of Physics, College of Science and Engineering, Ritsumeikan University
Noji-higashi 1-1-1, Kusatsu 525-8577, Japan}
\author{Kensuke S. Ikeda}
\affiliation{College of Science and Engineering, Ritsumeikan University
Noji-higashi 1-1-1, Kusatsu 525-8577, Japan}

\date{\today}
\begin{abstract}
Dynamical localization phenomena of  monochromatically
perturbed standard map (SM) and Anderson map (AM),
which are both identified with a two-dimensional disordered system 
under suitable conditions, are investigated by the numerical 
wavepacket propagation. Some phenomenological formula 
of the dynamical localization length 
valid for wide range of control parameters are proposed for both SM and AM. 
For SM the formula completely agree with the experimentally established formula, 
and for AM the presence of a new regime of localization is confirmed.
These formula can be derived by the self-consistent mean-field theory 
of Anderson localization on the basis of a new hypothesis for cut-off length.
Transient diffusion in the large limit of the localization length is also discussed.
\end{abstract}



\maketitle



\def\bsc{\begin{screen}}
\def\esc{\end{screen}}
\def\bit{\begin{itemize}}
\def\eit{\end{itemize}}

\def\red{\textcolor{red}}
\def\blue{\textcolor{blue}}
\def\green{\textcolor{green}}


\def\tr#1{\mathord{\mathopen{{\vphantom{#1}}^t}#1}}

\def\ni{\noindent}
\def\nn{\nonumber}
\def\bH{\begin{Huge}}
\def\eH{\end{Huge}}
\def\bL{\begin{Large}}
\def\eL{\end{Large}}
\def\bl{\begin{large}}
\def\el{\end{large}}
\def\beq{\begin{eqnarray}}
\def\eeq{\end{eqnarray}}

\def\eps{\epsilon}
\def\th{\theta}
\def\del{\delta}
\def\omg{\omega}

\def\e{{\rm e}}
\def\exp{{\rm exp}}
\def\arg{{\rm arg}}
\def\Im{{\rm Im}}
\def\Re{{\rm Re}}

\def\sup{\supset}
\def\sub{\subset}
\def\a{\cap}
\def\u{\cup}
\def\bks{\backslash}

\def\ovl{\overline}
\def\unl{\underline}

\def\rar{\rightarrow}
\def\Rar{\Rightarrow}
\def\lar{\leftarrow}
\def\Lar{\Leftarrow}
\def\bar{\leftrightarrow}
\def\Bar{\Leftrightarrow}

\def\pr{\partial}

\def\Bstar{\bL $\star$ \eL}

\def\>{\rangle} 
\def\<{\langle} 
\def\RR {\rangle\!\rangle} 
\def\LL {\langle\!\langle} 
\def\const{{\rm const.}}


\def\etath{\eta_{th}}
\def\irrev{{\mathcal R}}
\def\e{{\rm e}}
\def\noise{n}
\def\hatA{\hat{A}}
\def\hatB{\hat{B}}
\def\hatC{\hat{C}}
\def\hatJ{\hat{J}}
\def\hatI{\hat{I}}
\def\hatP{\hat{P}}
\def\hatQ{\hat{Q}}
\def\hatU{\hat{U}}
\def\hatW{\hat{W}}
\def\hatX{\hat{X}}
\def\hatY{\hat{Y}}
\def\hatV{\hat{V}}
\def\hatt{\hat{t}}
\def\hatw{\hat{w}}

\def\hatp{\hat{p}}
\def\hatq{\hat{q}}
\def\hatU{\hat{U}}
\def\hatn{\hat{n}}

\def\hatphi{\hat{\phi}}
\def\hattheta{\hat{\theta}}

\def\iset{\mathcal{A}}
\def\fset{\mathcal{B}}
\def\pr{\partial}
\def\traj{\ell}
\def\eps{\epsilon}

\def\DZ{D^{(0)}}
\def\Dcl{D_{cl}}

\def\noise{n}
\def\hatP{\hat{P}}

\def\U{U_{\rm cls}}
\def\P{P_{{\rm cls},\eta}}
\def\traj{\ell}
\def\cc{\cdot}


\section{Introduction} 
In one-dimensional quantum systems, 
strong localization phenomena have been observed 
due to large quantum interference effect
when disorder exists in the system.
A quite similar localization phenomenon occurs in 
classically chaotic dynamical systems which exhibit 
chaotic diffusion in the classical limit.
A typical example of the former is 
one-dimensional disordered systems (1DDS) \cite{ishii73,lifshiz88}, 
and the latter one is the quantum standard map (SM) \cite{casati79}. 
It has been shown that the localization of the wavepacket can be delocalized 
by applying dynamical perturbation composed of a few number 
of coherent modes
\cite{casati89,borgonovi97,chabe08,wang09,lemarie10,tian11,lopez12,lopez13}. 
If the number of the modes is more than two, the delocalization
takes place through a localization-delocalization transition (LDT) accompanied by
remarkable critical phenomena as the perturbation strength is increased.
It has been explored in detail the transition process from the localized phase 
for mode number more than two. 

In the previous paper \cite{yamada15}, we also investigated 
quantum diffusion of an initially localized wavepacket 
in the polychromatically perturbed Anderson map (AM) which is 
a time-discretized version of the Anderson model,  
 in comparison with  
the SM  driven by the same polychromatic perturbation.
However, the nature of quantum diffusion exhibited by the {\it monochromatically perturbed} 
AM and SM, which has been supposed to be localized, have not still been well-investigated, 
except for early stage studies on SM \cite{shepe83}.

Experimentally, Manai {\it et al.} observed the critical phenomenon of the 
LDT for Cesium atoms in optical 
lattice \cite{chabe08}, which is an experimental
implementation of the perturbed SM, and the observed results were 
successfully interpreted as a three-dimensional LDT based on the 
equivalence between SM and multi-dimensional disordered lattice 
by the so-called Maryland transform \cite{grempel84}.
Their results are also interpreted by the self-consistent theory (SCT) of 
the weak localization in three dimensional disordered system (3DDS) 
\cite{casati89,borgonovi97,chabe08,wang09,lemarie10,tian11,lopez12,lopez13}.  
Further, they recently observed the localization phenomenon in the SM 
driven by coherent monochromatic perturbation \cite{manai15}. This work 
is a very important experimental contribution in the sense that it first 
succeeded in realizing the two-dimensional disordered system (2DDS) 
as a monochromatically perturbed SM in the optical lattice.
In order to confirm the presence of localization, very long time-scale data 
must be examined, which is very difficult in real experiment but is much 
easier in numerical simulation. 
After the early report suggesting the presence of localization \cite{shepe83}, 
there has been no work of numerical simulation for the monochromatically perturbed SM.
The experimental results should be examined by reliable numerical 
simulation taking sufficiently long time steps, which will be done 
in the present paper.
We note also that there have been several studies on the localization of 
the copuled SM, which can be identified with the 2DDS by 
the Maryland transformation \cite{doron88,ballentine09}, 
but no definite quantitative results has been estabilished, because long 
time-scale simulation of coupled rotors is much more difficult than
the monochromatically perturbed SM.

There are two main purposes in the present paper. One is to explore 
systematically the localization characteristics of the monochromatically 
perturbed SM with the numerical and theoretical methods.
We focus our investigation on the {\it quantum regime} in which
the coupling strength is smaller than a characteristic value decided 
by the Planck constant, and reexamine the validity of the experimentally 
observed result of the Manai {\it et al.} in a wide parameter range of the 
quantum regime.
Our results are interpreted by the SCT of the localization based
on a newly proposed hypothesis.

Another purpose is to report the characteristics of monotonically perturbed
AM in comparison with the perturbed SM mentioned above,  
whose localization property has not been fully investigated.
The AM is close to the original model of random lattice proposed by
Anderson
 in a sense that randomness is explicitly included,
 and has its own physical origin quite different from the SM.
Note that there have already been some publications for numerical results
of AM \cite{yamada04,yamada12,yamada10}, 
and the presence of localization phenomenon for unperturbed
and monochromatically perturbed AM was stressed,   
but the present paper is the first detailed quantitative exploration 
of the dynamical localization length and the scaling properties for 
the monochromatically perturbed AM.
The parameter dependence of the localization length 
on the disorder strength and perturbation strength
are given numerically, and it is theoretically interpreted based 
on the SCT of the localization.

Our main concern is whether or not the above mentioned two models with 
quite different physical origin share common features of the dynamical 
localization phenomenon.
The outline of the paper is as follows.
In the next section, we introduce model systems, 
monochromatically perturbed SM and AM,
examined in the present paper.
The numerical results of scaling properties of the 
localization length in the perturbed 
SM and AM are given in Sects. \ref{sec:SM} and \ref{sec:AM}, respectively, and some 
empirical formula representing the localization characteristic are proposed. 
In particular, the existence of two different regimes of the localization is confirmed for AM. 
In Sect. \ref{sec:MarylandSCT}, these formula are consistently  
derived by the SCT of Anderson localization for anisotropic 
2DDS \cite{vollhard80,wolfle10} by 
introducing some hypothesis for the characteristic length as 
a cut-off in the self-consistent equation.
In addition, the relation between SM and AM is made clear by 
the Maryland transform \cite{grempel84}.
In Sect. \ref{sec:time-dep-Dt}, we discuss the characteristics of
diffusion for both models observed transiently on the way to
the final localization. The existence of the semiclassical regime 
beyond the quantum regime is emphasized for SM.
The last section is devoted to summary and discussion.
In appendixes, we give some complementary numerical results and a 
simple derivation of the Maryland transform
 in  Sect. \ref{sec:MarylandSCT}.

\section{Models}
\label{sec:model}
The model Hamiltonian of the periodically kicked system
driven by dynamical perturbation with different frequency $\omega_1$
from those of the kick is 
\beq
 H(\hatp,\hatq,t) 
=T(\hatp) + V(\hatq) \left\{ 1+\eps \cos(\omega_1 t +\varphi_0) \right\} \delta_t, 
\label{eq:H-time}
\eeq
where 
\beq
\delta_t =\sum_{m=-\infty}^{\infty}\delta(t-m\tau) 
 = \frac{1}{\tau}\sum_{m=-\infty}^{\infty}\cos(\frac{2\pi}{\tau}mt).
\label{eq:delta}
\eeq
Thus the system is kicked by
the periodic delta-functional force with the period $\tau$.
Here $T(\hatp)$ and $V(\hatq)$ represent translational kinetic energy and potential 
energy, respectively. And $\hatp$ and $\hatq$ are momentum and positional operators
of the kicked system, respectively.
The evolution for the single step between the time interval $[s\tau,(s+1)\tau]$ 
is represented by the unitary operator
\beq
 U(s\tau,\varphi_0) 
=\e^{-iT(\hatp)\tau/\hbar} 
\e^{-iV(\hatq)\left\{ 1+\eps \cos(\omega_1 \tau s +\varphi_0)\right\}/\hbar },
\label{eq:Ut}
\eeq
which depends explicitly upon the step $s \in {\Bbb Z}$.
The sinusoidal periodic perturbation is characterized by the frequency $\omega_1$
incommensurate with the kick frequency $2\pi/\tau$ and the strength $\eps$.
We can take the effect of the periodic force into account by introducing
an additional linear oscillator $\omega_1 \hatJ$ acting as the periodic force 
to the kicked system, which we call $J-$oscillator hereafter.
$\varphi_0$ is an initial phase of the  oscillation.
Then,  instead of the Hamiltonian of Eq.(\ref{eq:H-time}), 
we consider the Hamiltonian of two degrees of freedom
\beq
 H_{tot}(\hatp,\hatq,\hatJ,\hatphi,t) 
= T(\hatp) + \omega_1 \hatJ +V(\hatq) (1+\eps \cos \hatphi)\delta_t,
\label{eq:H-tot}
\eeq
where  $\hatphi=\phi, \hatJ=-i\hbar d/d\phi$ are the angle-action operators in the
angle representation,  and the angle variable is defined in the section $[0,2\pi]$.
The corresponding unitary evolution operator for the kick period $\tau$ does no longer depends upon 
the step $s$, and is an ``autonomous'' evolution operator 
\beq
 U_{tot} = \e^{-i\omega_1 \hatJ\tau/\hbar}\e^{-iT(\hatp)\tau/\hbar}\e^{-iV(\hatq)(1+\eps\cos\hatphi)/\hbar}.
\label{eq:Utot}
\eeq
One can show the relation between the autonomous evolution operator Eq.(\ref{eq:Utot}) and 
the non-autonomous evolution operator of Eq.(\ref{eq:Ut}) as 
\beq
\nn U_{tot}^s &=& 
{\cal T}\e^{-i\int_0^sH_{tot}(s')ds'/\hbar}\\
\nn  &=& \e^{-i\omega_1 J\tau s/\hbar}  
         {\cal T}\e^{-i \int_0^s dt^{'} \left[ T(\hatp)\tau + V(\hatq)\left\{ 1+
\eps \cos(\omega_1 t^{'} +\varphi_0)\right\}  \right] /\hbar}\\ 
 &=& \e^{-i\omega_1 J\tau s/\hbar}U(s\tau,\varphi_0)
U((s-1)\tau,\varphi_0)...U(\varphi_0).
\label{eq:Utot-t}
\eeq
where ${\cal T}$ is the time ordering operator.  Suppose that we take the action 
eigenstate, for example, $|J=0\>$ as the initial state
of the $J-$oscillator. It is represented by the Fourier sum over the angle eigenstates as 
$|J=0\> = \frac{1}{\sqrt{J}} \sum_{j}^J|\phi_j\>$, where $\phi_j =2\pi j/J$. 
Then, applying Eq.(\ref{eq:Ut}), the wavepacket propagation by $U_{tot}$ launched from the 
state $|J=0\>\otimes|\Psi_0\>$, where $|\Psi_0\>$ is an initial state of the kicked oscillator, 
is achieved by applying the periodically perturbed evolution operator $U(t\tau,\varphi_0)$ of Eq.(\ref{eq:Ut}) 
to the initial state $|\phi_0\> \otimes |\Psi_0\>$ and next summing over $\phi_0$. Summation over 
$\varphi_0$ can be replaced very well by the ensemble average over randomly chosen $\varphi_0$ \cite{memo1}.
We can thus use the representation of Eq.(\ref{eq:Ut}) for the numerical wavepacket propagation. But in
theoretical considerations we often return to the autonomous representation 
of Eq.(\ref{eq:Utot}). 

In the present paper we set $T(\hat{p})=\hat{p}^2/2$, $V(\hat{q})=K\cos \hat{q}$ for SM, and 
$T(\hat{p})=2\cos(\hat{p}/\hbar)=(\e^{\pr/\pr q}+\e^{-\pr/\pr q})$ (hopping between nearest neighbour sites), 
 $V(\hat{q})=Wv(\hat{q})=W\sum_n\delta(q-n) v_q|q\> \<q|$ for AM, respectively, 
where on-site potential $v_n$ takes random value uniformly distributed 
over the range $[-1, 1]$ 
and $W$ denotes the disorder strength. 

In the autonomous representation, the Heisenberg equation (classical equation) of 
motion describing the monochromatically perturbed SM is 
\beq
\left \{
 \begin{array}{llll}
p_{s+1}-p_s &=& K\sin q_{s}(1+\eps\cos\phi_s), \\
q_{s+1}-q_s &=& p_{s+1}\tau, \\
J_{s+1}-J_s &=& K\eps\cos q_{s} \sin\phi_s , \\
\phi_{s+1}-\phi_{s} &=& \omega_1\tau. 
 \end{array}
 \right.
\label{eq:SM-cl}
\eeq
where the Heisenberg operator is defined by $X_s\equiv U^{-s}XU^s$.
The set of equation for the monochromatically perturbed AM 
can be also obtained formally by the same way, but we should note that they have no
the classical counterpart.

Let us consider the relation between the discretized system
and the time-continuous system using unperturbed cases ($\eps=0$) for simplicity.
We take the symmetrized unit of time section $[s\tau-\tau/2,s\tau+\tau/2]$ instead of 
$[s\tau,(s+1)\tau]$, then we have the symmetrized form of the single step evolution 
operator $U_{SY} = \e^{-i\tau T/2\hbar}\e^{-iV/\hbar}\e^{-i\tau T/2\hbar}$ instead of Eq.(\ref{eq:Ut}).  
It is well-known that $U_{SY}$ approximates $\e^{-i\tau(T+V/\tau)/\hbar}$ 
up to the correction of $O((\tau/\hbar)^3)$, which is the lowest order 
Baker-Hausdorff-Campbell expansion 
of incommutable operator product. Hence, if we make $\tau/\hbar\ll 1$ 
keeping $V/\tau\sim O(1)$, the time evolution by $U_{SY}$ is closely approximated by that of the time-continuous Hamiltonian
$T+V/\tau$, and the AM agrees with the Anderson model of 1DDS, whereas the 
SM becomes the gravitational pendulum. 
(The condition $\tau/\hbar\ll 1$ is relaxed to $\tau\ll1$ if the system has 
the classical limit.)  We hereafter choose the period as $\tau=1$ 
throughout the present paper. 


We explain here the numerical wavepacket propagation based upon 
the nonautonomous unitary evolution operator 
Eq.(\ref{eq:Ut}) in which $\varphi$ is taken as classical random number. 
The localization phenomena take place in the momentum ($p$) space for SM and
in the position ($q$) space for AM, respectively. The momentum space of
SM and the position space of AM are spanned by the momentum or position eigenfunctions
commonly denoted by $|n\>$, where $n \in {\Bbb Z}$, and the momentum and position 
are quantized as $p=n\hbar$ for SM and $q=n$ for AM. Further the periodic boundary condition is 
imposed on the wavefunction $|\Psi\>$ as  $\<n+N|\Psi\>=\<n|\Psi\>$ 
for momentum (SM) and position (AM) 
representations, and so the integer value $n$ is bounded as $-N/2\leq n\leq N/2$ for 
a very large positive integer $N$.

Let the wavepacket at the time $t$ be 
\beq
|\Psi_s\>= U(s,\varphi_0)U((s-1),\varphi_0)....U(\varphi_0)|\Psi_0\>
\eeq
starting with the localized state $|\Psi_0\>=|n_0\>$ in the momentum (SM) or
in the position (AM) space. We monitor the time-dependent mean square displacement (MSD),  
$m_{2}(t)= \<\sum_{n=-\infty}^{\infty} (n-n_0)^2  |u(n,t)|^2\>_\Omega$ for the propagating 
wavepacket $u(n,t)=\<n|\Psi_t\>$, where
$\<\dots\>_\Omega$ denotes the ensemble average over initial condition $n_0$ for SM 
and different random configuration of $v(n)$ for AM, respectively.
In addition, the average over $\varphi_0$ should be taken, but the $\varphi_0$ dependence
of the MSD is much weaker compared with the dependency upon $n_0$ (for SM) and
sample of $v(n)$ (for AM), and the averaging is ignored if unnecessary.

\begin{figure}[htbp]
\begin{center}
\includegraphics[width=9cm]{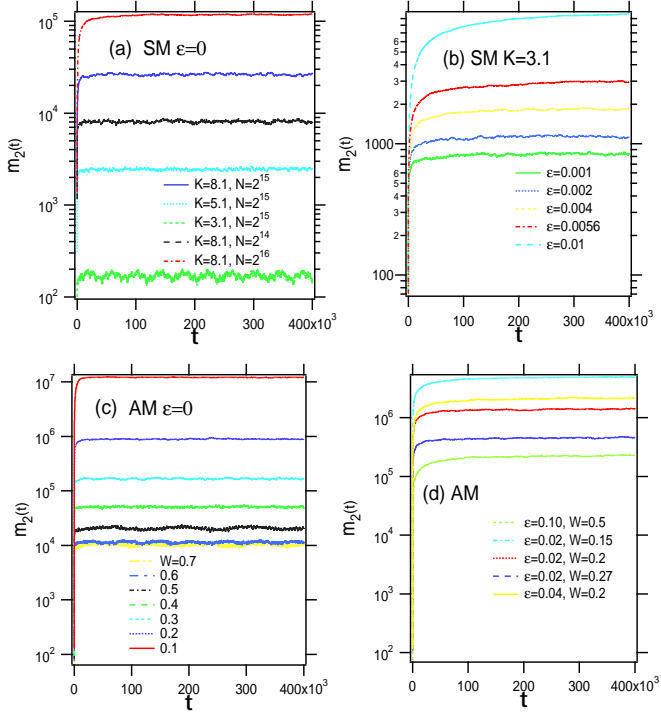}
\caption{(Color online)
The time-dependence of the MSD of SM and AM.
(a)Unperturbed SM for $K=3.1, 5.1, 8.1$ with $\hbar=\frac{2\pi 1248}{2^{15}}$,
and for $K=8.1$ with 
$\hbar=\frac{2\pi 1248}{2^{14}}$, $\frac{2\pi 1248}{2^{16}}$. 
(b)Monochromatically perturbed SM for $K=3.1$ and  $\hbar=0.12$
with $\eps=0.0001 \sim 0.1$ from below.
(c)Unperturbed AM with $W=0.1\sim 0.6$ from below.
(d)Monochromatically perturbed AM with some combinations of $\eps$ and $W$.
Note that the horizontal axes are in logarithmic scale.
The system and ensemble sizes are $N=2^{15}-2^{16}$ and $10-50$, respectively,
thorough this paper.
}
\label{fig:AMSM-fig1}
\end{center}
\end{figure}

In this paper,  we compute the localization length  (LL) of the dynamical localization, 
 $p_\xi=\sqrt{m_{2}(\infty)}$ for SM and 
$\xi=\sqrt{m_{2}(\infty)}$ for AM,  after numerically calculating the MSD for long-time,  
where $m_{2}(\infty)$ is numerically saturated MSD.
In fact, Fig. \ref{fig:AMSM-fig1} shows the time-dependence of MSD for some unperturbed 
and perturbed cases in SM and AM. 
It is found that the growth of time-dependence 
is saturated and the LL becomes larger values 
as the perturbation strength becomes larger.

Note that it is difficult to get the accurate LL 
as the perturbation strength increases 
for cases with the larger $K/\hbar$ in SM  and smaller $W$ in AM
 because of explosive increase of MSD.
Then we also use the time-dependent diffusion coefficients
to characterize the transient behaviour before reaching the LL
as will be discussed in Sect.\ref{sec:time-dep-Dt}.

\section{Localization phenomena in monochromatically perturbed quantum SM}
\label{sec:SM}
\def\const{\rm const.}
\def\k{\kappa}
In this section we show the localization characteristics 
of the monochromatically perturbed SM
with changing the three parameters $K$, $\hbar$ and $\eps$ 
in a wide range. 

As was already demonstrated in the experiments by the Manai {\it et al.}, 
a remarkable feature of the SM with monochromatic  perturbation
is a definite exponential growth of the localization length 
with respect to the perturbation strength $\eps$ \cite{manai15}, namely,  
\beq
\label{eq:SM1}
 p_\xi = D \exp \{ \eps A \},
\eeq 
where  the constants $A$, $D$ are determined by $K$ and $\hbar$.
The experiment of the Manai {\it et al.} was done for $\hbar$ greater 
than unity, i.e. in a strong quantum regime. 
On the other hand, it has been numerically and experimentally 
observed that, if $\hbar$ is small enough, classical diffusion of 
coupled SMs, which can be identified with 2DDS, 
is restored over a long time scale.
Problems related to the classical diffusion will be discussed in Sect.\ref{sec:time-dep-Dt}.

In this section, we focus on the results of numerical experiments 
in the "quantum regime" where $\eps$ is 
smaller than a certain characteristic
value dependent upon $\hbar$, which will be discussed later.

\begin{figure}[htbp]
\begin{center}
\includegraphics[width=8.0cm]{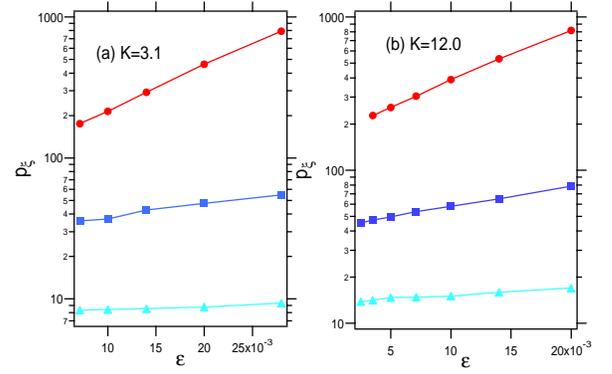}
\caption{(Color online)
Localization length $p_\xi$ as a function of the relatively small 
perturbation strength $\eps$ in the quantum regime
($\eps=1\times 10^{-3} \sim 30 \times 10^{-3}$).　
(a)$K=3.1$ for $\hbar=\frac{2\pi 435}{2^{12}}$, 
$\frac{2\pi 435}{2^{13}}$, $\frac{2\pi 435}{2^{14}}$ from below.
(b)$K=12.0$ for $\frac{2\pi 1741}{2^{12}}$, 
$\frac{2\pi 1741}{2^{13}}$, $\frac{2\pi 1741}{2^{14}}$ from below.
Note that the horizontal axes are in logarithmic scale.
}
\label{fig:SM-eps}
\end{center}
\end{figure}

Figure \ref{fig:SM-eps} shows $\eps-$dependence of the localization length 
$p_\xi$ for some $\hbar$'s and $K$'s. All the plots tell that the expression 
of (\ref{eq:SM1}) works quite well. Validity of Eq.(\ref{eq:SM1})
was confirmed for all values of $K$ and $\hbar$ we examined.
In the following, we discuss the $K-$dependence and the $\hbar-$dependence
of  the coefficients $A$, $D$, by the intercept and the slope numerically 
determined by the semi-log plot of Fig.\ref{fig:SM-eps}.

\begin{figure}[htbp]
\begin{center}
\includegraphics[width=8.0cm]{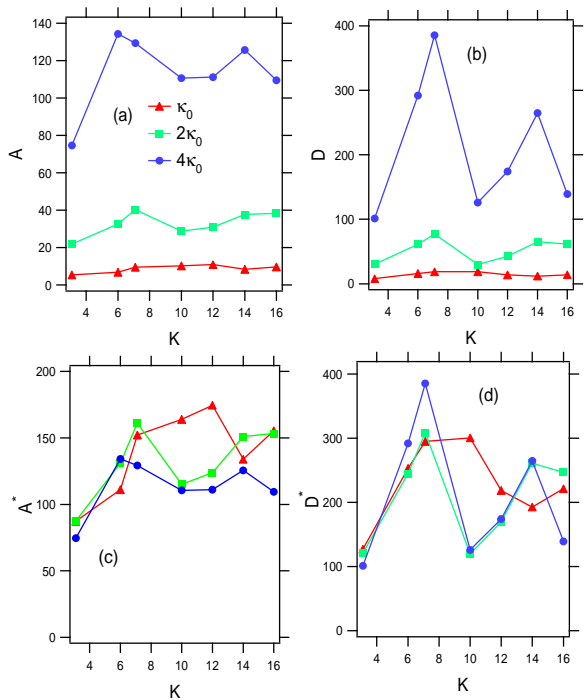}
\caption{(Color online)
$K-$dependence of 
the coefficients (a)$A$ and (b)$D$ of SM with a fixed $\k(\equiv K/\hbar)$.
The three cases for $\k=\k_0, 2\k_0,4\k_0$ ($\k_0=18$) are shown from below.
The scaled coefficients $A^*=A/(\k^2/\k_{max})$ and $D^*=D/(\k^2/\k_{max})$, 
where $\k_{max}=4\k_0$, are shown for $\k=\k0, 2\k0$ and, $4\k0$ in (c) and (d) 
of the lower panels, respectively. The three curves almost overlap.
}
\label{fig:SM-fig2}
\end{center}
\end{figure}

\begin{figure}[htbp]
\begin{center}
\includegraphics[width=8.0cm]{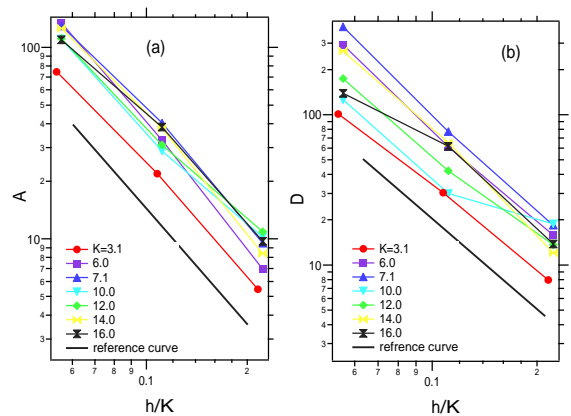}
\caption{(Color online)
(a)$(\hbar/K)-$dependence of the coefficients (a)$A$ and (b)$D$
for some $K$'s.
Three cases of  $\k=\k_0, 2\k_0, 4\k_0$ ($\k_0=18$) are shown 
in the  straight lines.
Note that the axes are in logarithmic scale.
The heavy line shows a straight line with slope of $-2$
corresponding to the $(\hbar/K)^{-2}$.
}
\label{fig:SM-fig3}
\end{center}
\end{figure}

First, we show the variation of the coefficient $A$ and coefficient $D$
for the change of $K$ with fixing the ratio $K/\hbar \equiv \k$.
It is shown in Fig.\ref{fig:SM-fig2}(a) 
for the three values $\k=\k_0,\k_0/2,\k_0/4$, where $\k_0=18$.
Obviously, it turns out that the variation in the value of the coefficient $A$ 
maintains a nearly constant value when the ratio $\k$ is constant 
if some irregular variation is ignored. These facts means that $A$ and $D$,
which is a function of $\k$ and $K$, should be the function of $\k$ alone.
Next, we change $\hbar$ for various fixed values of $K$, 
which is shown in Fig.\ref{fig:SM-fig3}(a).
Apparently, for all values of $K$ the dependence $A \propto \hbar^{- 2}$ is observed.
Since $A$ depends only on $\k$, the following relation should hold
\beq
\label{eq:SM2}
A  = {\rm const.} \left(\frac{K}{\hbar}\right)^2 \propto \k^2.
\eeq
Actually, the prediction is confirmed by the fact that the scaled coefficient $A/\k^2$ 
is almost constant for different $K$ and $\k$ (and so $K$ and $\hbar$) as is shown 
in Fig.\ref{fig:SM-fig2}(c).

On the other hand, the  $(K/\hbar)-$dependence 
of another coefficient $D$ is shown in Fig.\ref{fig:SM-fig2}(b) and Fig.\ref{fig:SM-fig3}(b).
It is expected that the coefficients $D$ and $A$ should show a similar behaviour
except for numerical prefactors, i.e. $D \sim A$. However, the $K-$dependence of 
the coefficient $D$ is less definite and it is often accompanied by some 
irregular fluctuations. This fluctuation becomes more pronounced as the $\k$ is 
larger, in other words, the smaller $\hbar$ enhances fluctuation,  
as recognized from Fig.\ref{fig:SM-fig3}(b).
This phenomenon is caused by the so-called acceleration modes 
which are peculiar to the classical dynamics of SM.
Actually, in the values of $K=2n\pi$ where the classical 
acceleration mode exists, the classical diffusion coefficient $D_{cl}$ 
increases explosively, 
and also reflects the localization length of the quantum system
as shown in the peak around $K=7$.
Increasing perturbation strength  $\eps$ reduces the effect of acceleration mode.
The acceleration modes existing with zero measure
in the classical system plays very complicated roles, but it is not essential 
to the discussion of the quantum localization phenomenon, 
so we will not discuss it in detail in this paper.
If we ignore such a fluctuation, the dependence of the coefficient $D$ 
on $K$ and $\hbar$ in Fig.\ref{fig:SM-fig2}(b) and Fig.\ref{fig:SM-fig3}(b) 
exhibits very similar behaviour
to the coefficient $A$ and so we conclude that
\beq
\label{eq:SM2D}
   D  = {\rm const.} \left(\frac{K}{\hbar}\right)^2 \propto \k^2.
\eeq
This prediction is confirmed also by the plot of the scaled coefficient $D/\k^2$ 
in Fig.\ref{fig:SM-fig2}(d) in a way parallel to Fig.\ref{fig:SM-fig2}(c).

To summarize the facts presented so far, 
localization length of monochromatically perturbed SM is represented by,
\beq
\label{eq:SM3}
  p_\xi \propto \left(\frac{K}{\hbar}\right)^2
 \exp \left[ \const \eps \left(\frac{K}{\hbar}\right)^2 \right]
\eeq
in the quantum regime. 
Note that in a limit of $\eps\to 0$, 
$p_\xi$ becomes the localization length of the unperturbed SM 
$p_\xi \propto (\frac{K}{\hbar})^2$ first proposed 
by Casati {\it et al} \cite{casati79}.

\begin{figure}[htbp]
\begin{center}
\includegraphics[width=6.0cm]{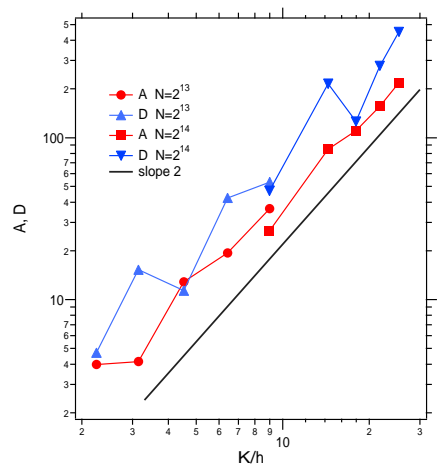}
\caption{(Color online)
$(K/\hbar)-$dependences of the coefficients $A$ and $D$
when varying the combination of $K$ and $\hbar$ variously.
Note that the axes are in logarithmic scale.
The heavy line shows a straight line with slope of $2$
corresponding to the $(\hbar/K)^{2}$.
}
\label{fig:SM-fig4}
\end{center}
\end{figure}

Finally, we plot the coefficients $A$ and $D$ 
in Fig.\ref{fig:SM-fig4} as a function of $K/\hbar$ 
fixing $\hbar$ at two significantly different values 
$\hbar=0.56$ and $3.1$ and changing $K$.
The coefficients $A$ and $D$ are not only proportional to 
$(K/\hbar)^2$, but the two curves of 
$A$ and of $D$ overlaps by extrapolation, 
which again establishes the results given by
Eqs.(\ref{eq:SM2}) and (\ref{eq:SM2D}).

The results obtained above agrees entirely with the experimental
results of the Manai {\it et al.}. They explained their results
by applying self-consistent mean field theory (SCT) 
to the monochromatically perturbed SM, and they obtained
$A \propto (\frac{K}{\hbar})^2$, but $D \propto\frac{K}{\hbar}$, which
is inconsistent with Eq.(\ref{eq:SM2D}).
A modified version of SCT of the localization naturally leading 
to the result Eq.(\ref{eq:SM3}) will be presented in Sect.\ref{subsec:SCT}.
The modification is essential for the theoretical prediction of
the localization length of AM discussed in the next section.

\section{Localization phenomena in monochromatically perturbed AM}
\label{sec:AM}
In this section,  we show the numerical results for the localization
characteristics of the monochromatically perturbed AM.
The scaling properties of LL is explored by varying
disorder strength $W$ and perturbation strength $\eps$.

\subsection{$W-$dependence}
\label{sec:AM_W}
It has been analytically found that in 1DDS the $W-$dependence of the
LL of eigenstates behaves like $W^{-2}$ for 
weak disorder limit $W<<1$ by perturbation theory \cite{kappus81,derrida84}
and it decreases obeying $1/\log W$ 
in the strong disorder limit $W>>1$ \cite{pichard86, kroha90}.
Therefore, if  the LL is almost same as the dynamical localization length, 
we can expect 
the time-dependent spread of the initially localized wavepacket is 
suppressed around the LL, and the  $W-$dependence of  the saturated MSD
behaves like $m_2(t)(=\xi_0^2) \sim W^{-4}$ in the weak disorder limit.

\begin{figure}[htbp]
\begin{center}
\includegraphics[width=8.0cm]{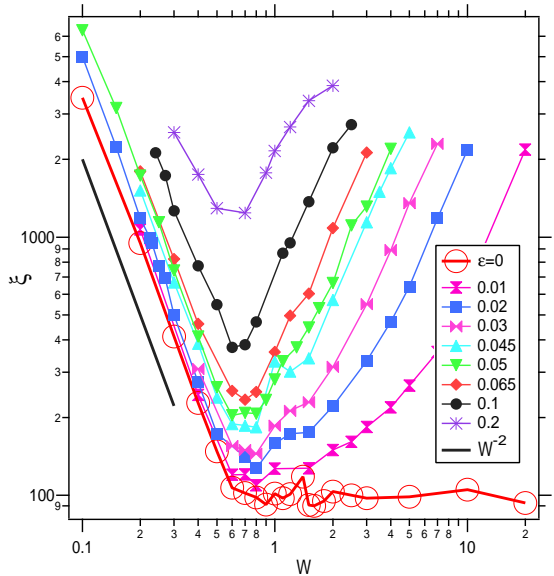}
\caption{(Color online)
Localization length of the monochromatically perturbed AM 
as a function of disorder strength $W$ for various perturbation strength $\eps$.
The unperturbed case ($\eps=0$) is denoted by a thick line with large circles.
Note that the axes are in the logarithmic scale.  
}
\label{fig:AM-fig2}
\end{center}
\end{figure}

Based on these facts, we investigate the localization properties of the wavepacket 
in the monochromatically perturbed AM.
Figure \ref{fig:AM-fig2} shows the $W-$dependence of the LL in the system 
for various perturbation strength $\eps$.
First, let us focus on the unperturbed case $(\eps=0)$ for which a typical situation 
of localization in AM is expected to occur.
It follows that the LL $\xi_0$ of the unperturbed case decreases like $W^{-2}$ in 
the weak disorder regime as expected, but the decrease ceases around a certain 
value denoted by $W^*$, that is, 
\beq
\xi_0 \simeq \left\{
\begin{array}{ll}
c_0W^{-2} & (W<W^*) \\
\xi_0^{*} & (W>W^*) 
\end{array}
\right.
\label{eq:1D_loclen}
\eeq
where $c_0$ is a constant and $\xi_0^*=\xi_0(W^{*})$.
The result is consistent with the perturbation theory  
only for the limit $W << 1$, as mentioned in Sect.\ref{sec:model}.
Its reason is reconsidered with the Maryland transform 
in the next section, 
but very intuitively we can explain the presence of the 
characteristic value $W^*$ above which the $\xi_0\sim W^{-2}$ behaviour 
changes to $\xi_0 \sim const$ by the periodic nature 
of the dynamical perturbation Eq.(\ref{eq:delta}) 
in Sect.\ref{sec:model}. 
Eq.(\ref{eq:H-time}) together with Eq.(\ref{eq:delta}) allows us 
to interpret the original Hamiltonian of AM (set $\eps=0$ for simplicity) as
the Anderson model Hamiltonian $T(\hatp) + V(\hatq)/\tau$ to which the dynamical
perturbation $(2/\tau)\sum_{m=1}^{\infty} \cos(2\pi mt/\tau)$ of the period $\tau$ is added.
The latter induces a transition between the localized eigenstates of the Anderson model 
if the typical energy width $W$ of the localized states exceeds the 
minimal quantum unit $\hbar 2\pi/\tau$ of the periodic perturbation.
Accordingly, it is expected that the $\delta-$function weakens the localization effect 
when $W$ exceeds the characteristic value   
\beq
   W^* \simeq \frac{2\pi}{\tau}\hbar.
\label{eq:W*}
\eeq
Taking $\hbar=1/8$, $\tau=1$, the above formula yields
$W^{*} \sim 0.8$ which is consistent with the characteristic value of
Fig.\ref{fig:AM-fig2} above which the monotonous decrease obeying 
the $W^{-2}-$law ceases. Indeed, we have confirmed the change of the 
value $W^*$ obeys Eq.(\ref{eq:W*}) by varying the period $\tau$.

The curves of the LL for various values of the perturbation strength $\eps$
is over-plotted in Fig.\ref{fig:AM-fig2}.
For $W<W^{*}$, the $W^{-2}$-dependence is stably maintained even for $\eps \neq 0$
but the LL increases with increase in the perturbation strength $\eps$
at least in the weak perturbation limit $\eps<<1$.
On the other hand, for $W>W^{*}$, the LL grows up
as the disorder strength $W$ increases if $\eps\neq 0$. 
In the next subsection, we look into the details of the $\eps-$dependence of the LL 
for the two regions, i.e., $W<W^{*}$ and $W>W^{*}$, to clarify their characteristics .

\subsection{$\eps-$dependence}
\begin{figure}[htbp]
\begin{center}
\includegraphics[width=9.0cm]{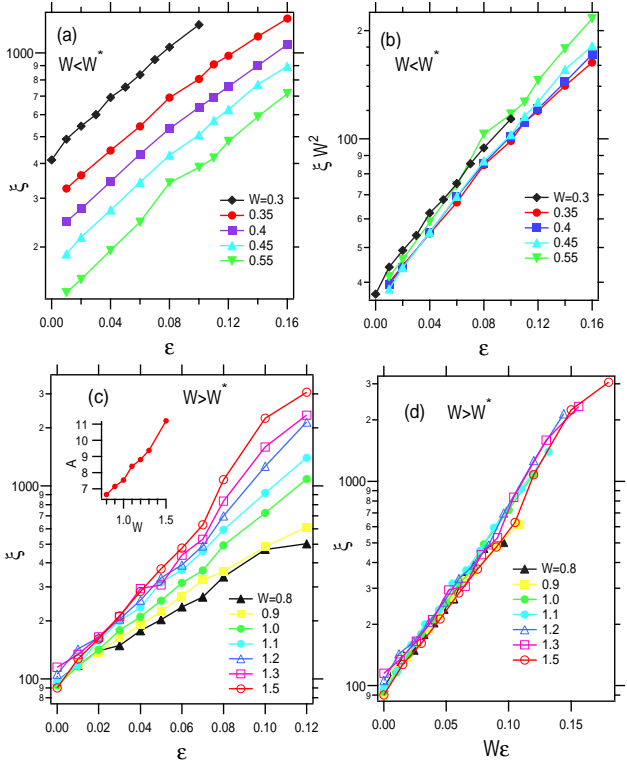}
\caption{(Color online)
Localization length of the monochromatically perturbed AM 
as a function of perturbation strength $\eps$ for 
(a) $W<W^*$, (c) $W>W^*$, where $W^*=0.8$.
(b)Plot of  $\xi W^2$ as a function of $\eps$ for  $W<W^*$.
(d)Plot of   $\xi$ as a function of $\eps W$ for  $W>W^*$.
Note that the all the vertical axes are in the logarithmic scale. 
The inset in panel (c) shows the plot of the coefficient $A$ as a function of $W$, 
which are estimated by linear fitting for data in the panel. 
}
\label{fig:AM-fig3}
\end{center}
\end{figure}

Figure \ref{fig:AM-fig3}(a) and (c) show the result of the $\eps-$dependence in the 
the monochromatically perturbed AM　for $W<W^*$ and  $W>W^*$, respectively.
It is obvious that the 
LL grows exponentially as 
the perturbation strength $\eps$ increases in the both cases. 
Therefore, in the same way as the case of SM the LL can be expressed as 
\beq
\xi \simeq D \exp \{ A \eps \}.
\eeq
The coefficient $D$ should be the LL at $\eps=0$ and so $D=\xi_0$,
whose characteristics are given as Eq.(\ref{eq:1D_loclen}).
The most interesting point is the $W-$dependence of the coefficient $A$ 
in the two characteristic regions, $W<W^*$ and $W>W^*$.
For $W < W^{*}$ the coefficient $A$ is almost independent of $W$, 
and $\xi W^2 \propto \xi/\xi_0=\xi/D$ as a function of  $\eps$ overlaps 
each other as shown in Fig.\ref{fig:AM-fig3}(b).
As a result,  we can obtain the relation  $\xi \sim c_0W^{-2} exp \{ c_1\eps \}$,
where $c_0$ and $c_1$ are certain constants.

On the other hand, in the region $W > W^{*}$ peculiar to AM,
there is a trend that the coefficient $A$ increases with $W$ 
as seen in Fig.\ref{fig:AM-fig3}(c).
The inset in panel (c)  shows the slope (coefficient $A$) of the data in the 
panel (c) determined by fitting  in the range of $0.01 <\eps < 0.08$ 
using  the method of least squares.
The $W-$dependence of  the slope increases almost linearly.
Indeed, all the plots of the LL as a function of $W\eps$ overlap
as shown in Fig.\ref{fig:AM-fig3}(d) for $W\eps <<1$. 
The same scaling behaviours have been observed for the cases
with different frequency $\omega_1$ 
as given in appendix \ref{app:different-frequency}. 

As a result, the LL $\xi$ of the monochromatically perturbed AM 
can be summarized as follows;
\beq
\xi \simeq \left\{
\begin{array}{ll}
c_0 W^{-2} exp \{ c_1\eps \} & (W<W^*) \\
\xi_0^{*} exp \{ c_2 W \eps \} & (W>W^*) ,
\end{array}
\right.
\label{eq:AM-last}
\eeq
where $c_0, c_1, c_2$ are numerical constants and $\xi_0^{*}$ is the saturated LL of
the unperturbed case.

The $W-$dependence of the coefficient $A$  is very different from 
those in SM although the exponential growth with respect to $\eps$
is common.

Beyond $W^*$ the transition between the localized states 
due to the dynamical perturbation play the role of stopping the decrease 
of LL as is exhibited by Eq.(\ref{eq:1D_loclen}). Recalling that the 
dynamical part  of perturbation potential  in AM is given by 
$\eps Wv(n)\cos(\omega_1t)$, it is interesting to see 
the fact that the perturbation amplitude $W\eps$ does not
influence the LL until $W$ exceeds $W^*$, which means that
the effect of dynamical perturbation fully works only after the 
transition channel opens.  Interpretation of Eq.(\ref{eq:AM-last}) 
by SCT of the localization will be presented in next section.

\section{Theoretical explanation}
\label{sec:MarylandSCT}
In this section, we first confirm the relationship among SM, AM and 2DDS by
the Maryland transform \cite{grempel84}.
Next, we give a theoretical explanation for the scaling properties obtained 
numerically in the last two sections is given based on the self-consistent 
mean-field theory (SCT) of the Anderson localization in 2DDS \cite{wolfle10}. 

\subsection{autonomous representation and Maryland transformation}
\label{subsec:Maryland}
We return to the two degrees of freedom unitary-evolution operator (\ref{eq:Utot}) which takes the monochromatic dynamical perturbation into account by the $J-$oscillator in an autonomous way.  
\beq
  \hatU_{tot}=\e^{-i\hatA}\e^{-i\hatB}\e^{-i\hatC},
\eeq
where 
\beq
    \e^{-i\hatA} &=& \e^{-\frac{i}{\hbar}[T(\hatp)+\omega_1 \hatJ]\tau},\\
    \e^{-i\hatB} &=& \e^{-\frac{i}{\hbar}\eps \hatV(q)\cos\phi\tau},\\
    \e^{-i\hatC} &=& \e^{-\frac{i}{\hbar}V(\hatq)\tau}.
\eeq
$\tau=1$ in this paper.
We consider an eigenvalue equation,
\beq
   \hatU_{tot}|u_1\> =\e^{-i\gamma}|u_1\>
\label{eq:eigen-value-problem}
\eeq
where $\gamma$ and $|u_1\>$ are the quasi-eigenvalue and quasi-eigenstate, 
respectively. This eigenvalue problem can be mapped into the tight-binding 
form by the Maryland transform, 
which provides with the foundation for
applying the analysis developed for the 2DDS to our systems. 
This formulation further gives rise to some remarks about our approach.

For the SM, the eigenvalue equation we take the representation using
eigenstate $|m\> (m\in{\Bbb Z})$ of momentum $\hatp$ and the action eigenstate 
$|j\> (j\in{\Bbb Z})$ of the $J$-oscillator as $u_1(m,j)=(\<m|\otimes \<j|)|u_1\>$.
Then by applying the Maryland transform, Eq.(\ref{eq:eigen-value-problem}) is transformed into 
the eigenvalue equation of the following two-dimensional 
lattice system (tight-binding model) with aperiodic and singular on-site potential
for newly defined eigenfunction $|u\>$ related to the original one $|u_1\>$ with
an appropriate transform shown in an appendix \ref{app:Maryland}.
\begin{widetext}
\beq
\label{eq:Maryland_SM}
\tan \left[ \frac{\hbar^2m^2/2+j\omega_1\hbar}{2\hbar}\tau-\frac{\gamma}{2} \right] u(m,j)+ 
 \sum_{m',j'}\<m,j|\hatt|m',j'\>u(m',j')=0
\eeq
where the transfer matrix element is
\beq
 \<m,j|\hatt|m',j'\> 
&=& \frac{1}{(2\pi)^2}\int_0^{2\pi} \int_0^{2\pi}  dq d\phi \e^{-i(m-m')q}\e^{i(j-j')\phi} 
\tan \left[\frac{K\cos q (1+ \eps \cos\phi)}{2\hbar}\tau  \right].
\label{eq:transfer-SM}
\eeq
\end{widetext}
This is the Maryland transformed eigenvalue equation including the additional degree 
of freedom contributing as the  monochromatic perturbation in the monochromatically 
perturbed SM. The details of the derivation is given in appendix \ref{app:Maryland}.
Of particular note is that in the semiclassical limit $\hbar \to 0$
the potential term become singular. Indeed, under the condition $|K\tau/\hbar|>\pi$
the transfer matrix element become the Fourier coefficient of a function having poles on
the real axis and the transfer matrix element do not decay as $|m-m'|\to \infty$
and so the analogy with the normal 2DDS is lost.

On the other hand, for the AM, we use the representation
 $u(n,j) (=(\<n|\otimes \<j|)|\ovl{u}\>)$
based on the eigenstates $|q=n>$ of the site  operator $\hat{n}$ 
and the eigenstates $|j\>$ of the operator $\hatJ$, and
the Maryland transformed eigenvalue equation becomes 
\begin{widetext}
\beq
\label{eq:Maryland_AM}
 \tan \left[ \frac{Wv_n+j\omega_1\hbar}{2\hbar}\tau-\frac{\gamma}{2}
 \right]u(n,j)+  \sum_{n',j'} \<n,j|\hatt|n',j'\>u(n',j')=0, 
\eeq
where 
\beq
\<n,j|\hatt|n',j'\> &=& \<n,j|i\frac{
e^{-i\eps Wv(\hatq)\cos\phi\tau/\hbar}-
e^{i2\cos(\hatp/\hbar)\tau/\hbar}
}
{e^{-i\eps Wv(\hatq)\cos\phi \tau/\hbar}+
e^{i2\cos(\hatp/\hbar)\tau/\hbar}
}|n',j'\>.
\eeq
\end{widetext}
In the case of $\eps \neq 0$, the evaluation of matrix elements is not easy
since the stochastic quantity $v_n$ is contained in addition to both operators  
$\hatq$ and $\hatp$.

If we take $\eps=0$ as the simplest case of the hopping term
in the transformed equation of AM, it becomes 
\beq
& & \<n,j|\hatt|n',j'\> \\
&=&\frac{1}{2\pi}\int_0^{2\pi}dp \e^{i(n-n')p}
\tan \left[\frac{\cos(p/\hbar)}{\hbar}\tau \right], 
\eeq
where $p=2\pi\hbar k/N$.
In the small $\tau$ limit the above equation results in an eigenvalue equation of 
the Anderson model with the nearest neighbouring hopping because $\tan(x) \simeq x$.
On the other hand as $W$ increases such that $\tau W/2\hbar$ exceeds $\pi/2$, 
the range of the 
on site potential of Eq.(\ref{eq:Maryland_AM}) covers the maximal range beyond which 
the distribution of the onsite potential do not change. This is an alternative
explanation for the saturation of the localization length beyond $W^*$, which has been
discussed in subsection \ref{sec:AM_W}.

Furthermore, from the Maryland transformed Eqs. (\ref{eq:Maryland_SM}) and (\ref{eq:Maryland_AM}), 
with $\eps=0$ the relationship between AM and SM can also be roughly estimated.
 In case of nearest neighbouring hopping for AM, the disorder strength increases with 
$W/\hbar$.
On the other hand,  the hopping strength increases with increase of $K/\hbar$  for $K/\hbar<<1$
in the case of SM. If the hopping strength is normalized to be unity, the disorder strength 
becomes proportional to  $\hbar/K$.
Accordingly, we can also see that the correspondence is roughly given as,
\beq
  \frac{W}{\hbar} \Leftrightarrow  \frac{\hbar}{K}.
\label{eq:KR}
\eeq

\def\DZ{D^{(0)}}
\subsection{Interpretation of the scaling properties based on self-consistent theory
of the localization}
\label{subsec:SCT}
Using the self-consistent theory of the mean-field approximation for the 
localization in the anisotropic 2DDS,  we interpret the scaling 
characteristics on the numerical results for SM obtained in Sect.\ref{sec:SM} 
and AM in Sect.\ref{sec:AM}, respectively.

Let $D_{\mu}(\omega)$ be the dynamical diffusion constant in the $\mu$ direction ($\mu=1,2$).
It is modified from the bare diffusion constant $\DZ_\mu$ due to the destructive
quantum interference induced by the backward scattering process
 of potential and is determined by the following self-consistent equation:   
\begin{eqnarray}
\label{eq:SCT-0}
  \frac{D_{\mu}(\omega)}{\DZ_{\mu}}
 = 1 - \frac{1}{\pi \rho} \frac{D_{\mu}(\omega)}{\DZ_{\mu}}
\sum_{q_1, q_2}\frac{1}{-i\omega+\sum_{\nu=1}^2D_{\nu}(\omega)q_\nu^2}.
\end{eqnarray}
The second term in the righthand side indicates the reduction by the quantum
interference effect. ($\rho$ is the density of states.)
In the localized phase, the $\omega-$dependent diffusion coefficient 
has a form $D_{\mu}(\omega) \propto -i\omega$, and 
is related to a scale of the  length $\xi(\omega)$ in the infinite system
as follows:
\begin{eqnarray}
\label{eq:SCT-0d}
 \xi_{\mu}(\omega)^2=D_{\mu}(\omega)/(-i\omega), 
\end{eqnarray}
which indeed becomes the localization length 
$\xi(\omega=0)$( or $p_{\xi}(\omega=0)$), 
in the limit of $\omega \to 0$.
Here the summation over the wavenumber $q_\mu$
is done up to the upper cut-off decided by the inverse of the characteristics
length $\ell_\mu$'s which are important parameters discussed below in detail. 
Then Eq.(\ref{eq:SCT-0}) in the $\mu-$direction  is rewritten by an integral form
\beq
& & \frac{\xi_{\mu}(\omega)^2}{\ell_{\mu}^2} (-i\omega t_\mu)  \nn \\
&=& 1 - \frac{\xi_{\mu}(\omega)^2t_\mu}{\ell_{\mu}^2}\frac{1}{\xi_1(\omega)\xi_2(\omega)}
\Xi \left[ \frac{\xi_1(\omega)}{\ell_1},\frac{\xi_2(\omega)}{\ell_2}  \right],
\label{eq:SCT-1}
\eeq
where $t_\mu=\ell_\mu^2/D_\mu^{(0)}$ means the localization time, and
\beq
& & \Xi\left[ \frac{\xi_1(\omega)}{\ell_1},\frac{\xi_2(\omega)}{\ell_2}\right] \nn \\
&=& \tilde{c} \int_0^{\xi_1(\omega)/\ell_1}\int_0^{\xi_2(\omega)/\ell_2} dQ_1 dQ_2
\frac{1}{1+Q_1^2+Q_2^2} ,
\label{eq:SCT-2}
\eeq
where $\tilde{c}$ is an appropriate numerical factor of $O(1)$.
Here, the characteristic length $\ell_\mu$ of the  integration range
is usually taken as the mean free path,  but in this paper we will propose different  characteristic length 
as shown below.

Taking the case of SM as an example, we show 
the difference of the length proposed in this paper (maximum distance) 
from the ordinary length (minimum distance) 
as the characteristic length  $\ell_\mu$.
Let the kicked system with the characteristic length $\ell_1$
be the main system and the $J-$oscillator 
with the characteristic length $\ell_2$ the subsystem.
The ordinary selection for $\ell_\mu$ is  the minimum distance
given as the hopping length $p(s+1)-p(s)=K\sin q_s$
 for a single step evolution  from Eq.(\ref{eq:SM-cl}).
The mean square values are
\beq
(\ell_1\hbar)^2 &=& K^2\< \sin^2q_s \>=K^2/2, \\
(\ell_2\hbar)^2 &=& K^2\eps^2\< \cos^2q_s \>\sin^2\phi_s=K^2\eps^2/4.
\eeq
Here $\<...\>$  indicates the quantum mechanical average with respect to the initial state. 
These correspond to the so called mean free path.
An  another candidate is the maximum distance reachable in an infinite time scale
represented  by the total hopping length 
$p(\infty)-p(0)=\lim_{s\to\infty}K \sum_{s'<s}\sin q_{s'}$.
Now we are considering the weak perturbation limit of $\eps$ in 
which the kicked system is decoupled from
the subsystem,  and maintains the diffusive motion  as
$m_2(s)=\<(p(s)-p(0))^2\>=D_{cl}s$,  
where $D_{cl}$ is the classical diffusion constant, until the localization time 
which should coincides with the number 
of states in the maximum length ,i.e. $\ell_1$ of the main system.
(This corresponds to the so-called Heisenberg time, and precisely 
a numerical factor must be multiplied but we ignore it.)

On the other hand, the $J-$oscillator  (the color degrees of freedom) 
also exhibits a passive diffusive motion up to 
\beq
\label{eq:equaltime}
 t_2 = t_1 = \ell_1,
\eeq
being driven by the same force as the main system.
(See Eq.(\ref{eq:SM-cl}).) 
From Eq.(\ref{eq:SM-cl}) the MSD's are expressed:
\beq
\label{eq:ell1ell0}
\lim_{s\to\infty}\<(p(s)-p(0))^2\> &=& \lim_{T\to \infty}\sum_{s \leq T}D_{1s}, \\
\lim_{s\to\infty}\<(J(s)-J(0))^2\> &=& \lim_{T\to \infty}\sum_{s\leq T}D_{2s},
\eeq
where the time-dependent diffusion constants $D_{1s}$ and $D_{2s}$ are defined by  
\beq
\label{eq:ell1ell1-1}
 D_{1s}& =& K^2[\<\sin^2q_s\>+{\rm Re}\sum_{s'<s}\<\sin q_{s'}\sin q_s\>]\, 
\eeq
\beq
\label{eq:ell1ell1-2}
D_{2s} &=& K^2\eps^2[\<\cos^2q_s\>\sin^2\phi_s  \nn \\
&+& {\rm Re}\sum_{s'<s}\<\cos q_{s'}\cos q_{s}\>\sin\phi_{s'} \sin\phi_{s}] .
\eeq
As the maximum diffusion lengths, $\ell_1$ and $\ell_2$, are proposed
\beq
\label{eq:ell1ell1d}
\ell_1^2\hbar^2 &=& \lim_{s\to\infty}\<(p(s)-p(0))^2\>, \\
\ell_2^2\hbar^2 &=& \lim_{s\to\infty}\<(J(s)-J(0))^2\>.
\eeq
The time-dependent diffusion constants are a given as
$D_{1s}=\DZ_{1}\hbar^2=D_{cl}$ and 
$D_{2s} \equiv \DZ_{2}\hbar^2 \sim \Dcl\eps^2/2$  until $t_1=t_2=\ell_1$, 
but both collapse to zero beyond it.  
Then Eqs.(\ref{eq:ell1ell0})-(\ref{eq:ell1ell1d}) lead to
\begin{eqnarray}
\label{eq:ell1ell2}
\ell_1^2 &=& \DZ_{1}\ell_1, \\
\ell_2^2 &=& \DZ_{2}\ell_1. 
\end{eqnarray}
This is a general relation applicable to the case of 
AM as will be used later.  If we suppose the Markovian limit that the 
autocorrelation function of the force term coming from the main system 
is given by $\<\cos q_{s}\cos q_{s'}\>
=\<\sin q_s\sin q_{s'}\>=\delta_{s,s'}/2$, then $\DZ_{1}=D_{cl}/\hbar^2=K^2/2\hbar^2$, 
 $\DZ_{2}=\eps^2\DZ_{1}/2$, and $\ell_{1,2}$ is given as
\begin{eqnarray}  
 \ell_1 &=& \DZ_{1}=\frac{D_{cl}}{\hbar^2},  \nn \\
 \ell_2 &=& \frac{\eps}{\sqrt{2}}\DZ_{1}.
\end{eqnarray}
As the fundamental distance $\ell_1$, $\ell_2$, we use these ``maximal distance'' rather than
the ``minimal distance'' taken by the Manai {\it et al.} \cite{manai15}.
We further remark that one can easily check the one-dimensional version of Eq.(\ref{eq:SCT-1}) 
can give the localization length of the standard map $\frac{D_{cl}}{\hbar^2}$ only by assuming
$\ell_1 = \DZ_{1}=\frac{D_{cl}}{\hbar^2}$.

Under the above setting, Eq.(\ref{eq:SCT-1}) tells that  the factor
$\frac{\xi_{\mu}(\omega)^2}{\ell_{\mu}^2}t_\mu$ is independent of $\mu$, which means that
\beq
\label{xi_ell}
\frac{\xi_{1}(\omega)}{\ell_{1}}=\frac{\xi_{2}(\omega)}{\ell_{2}}
\eeq
because $t_{\mu}=t_1$ from Eq.(\ref{eq:equaltime}). 
Carrying out the integral in the r.h.s of Eq.(\ref{eq:SCT-1}) for $\mu=1$, using the
above relation, one has
\beq
& & \frac{\xi_{1}(\omega)^2}{\ell_{1}^2} (-i\omega\ell_1)  \nn \\
&=& 1 - \frac{\tilde{c}}{\ell_2}\log \left[ 1+\frac{\xi_1(\omega)^2}{\ell_1^2} \right].
\label{eq:SCT-3}
\eeq

Taking a limit $\omega \to 0$ and organizing the expressions, 
the localization length $\xi_1(0)$ becomes
\beq
 \xi_1(\omega=0) \sim \ell_1\e^{\ell_2/2\tilde{c}},
\label{eq:loc-len}
\eeq
where $\ell_1=\frac{D_{cl}}{\hbar^2}=\frac{K^2}{2\hbar^2}$ and 
$\ell_2=\frac{\eps D_{cl}}{\sqrt{2}\hbar^2}=\eps\frac{K^2}{2^{3/2}\hbar^2}$
are two selected characteristic lengths, and $\tilde{c}$ is a suitable constant.
This corresponds to the localization length in SM under 
the monochromatic perturbation in the previous sections.
We remark that in the case of typical isotropic 2DDS 
the characteristic length is 
$\ell_1=\ell_2(=\ell)$ can be identified with the mean free path $\ell_{mfp}$
and Eq.(\ref{eq:loc-len}) yields  the well-known result 
$\xi_{2dds}\sim \ell_{mfp} \e^{\pi \ell_{mfp}/2}$.

We can straightforwardly apply the above analysis to 
the perturbed AM. In a similar way as in the case of SM, 
the diffusion length of the $J-$oscillator is obtained by
replacing the term $K\eps\cos q_s\sin\phi_s$ in Eq.(\ref{eq:SM-cl}) coming from the 
interaction potential $K\eps \cos q_s \cos \phi_s$ 
by the term $W\eps v(q_{s})\sin \phi_s$ 
in the interaction potential of the AM as, 
\beq
\ell_2^2\hbar^2 & \sim & W^2\eps^2 \sum_{s,s'}\<v(q_{s})v(q_{s'})\>\<\sin\phi_s\sin\phi_{s'}\>. 
\eeq
Since the diffusion time of the kicked system is given by  $\ell_1$, 
\beq
\ell_2^2\hbar^2  \sim   W^2\eps^2 \ell_1/2.
\eeq
Accordingly,  in the case of AM 
\beq
  \ell_1 
\simeq \left\{
\begin{array}{ll}
1/W^2 & (W<W^*) \\
1/W^{*2} & (W>W^*) .
\end{array}
\right.
\eeq
Therefore, 
\beq
  \ell_2 \sim \eps\sqrt{W^2 \ell_1/2}   
\simeq \left\{
\begin{array}{ll}
\eps/\sqrt{2} & (W<W^*) \\
\eps W/\sqrt{2} & (W>W^*) .
\end{array}
\right.
\eeq
It follows that when these are used for the expression (\ref{eq:loc-len}),
 results are consistent with that obtained by  the numerical calculation
in the previous sections.

In the SM, it also means that what the Manai {\it et al.} 
used to explain the experimental results could be derived directly 
from the theoretical considerations by our selection 
for the characteristic lengths as the cut-off of the integral of
SCT of the localization.
Our hypothesis Eq.(\ref{eq:ell1ell1d}) is more vital in SM dynamically 
perturbed by more than two colors, in which a localization-delocalization 
transition occurs \cite{lopez13}. Indeed, it predicts precise 
parameter dependence of the critical value of $\eps$ numerically 
observed \cite{yamada15}.

\section{
Diffusion Characteristics}
\label{sec:time-dep-Dt}
Up to the previous section, we investigated localization characteristics
in a relatively small perturbation regime in which the LL
can be decided numerically. The exponential growing rate of the localization 
length is enhanced by decreasing $\hbar$ or by increasing $W$ for SM and
AM, respectively. 
It is quite interesting to see the transient 
behaviour on the way to the final localization 
in the large limit of the exponentially enhanced LL.
In the case of AM, the region $W>W^*$ is focused on, because it is an essentially new 
region peculiar to the quantum map in which the quantum hopping is assisted 
by the kick perturbation.
On the other hand, in the case of SM the limit $\hbar\to 0$ is of interest because it is 
the semiclassical limit in which the LL is much enhanced, 
and classical chaotic diffusion may be observed at least in the transient process, 
as has been first examined in coupled SM \cite{adachi88}.

\begin{figure}[htbp]
\begin{center}
\includegraphics[width=9.0cm]{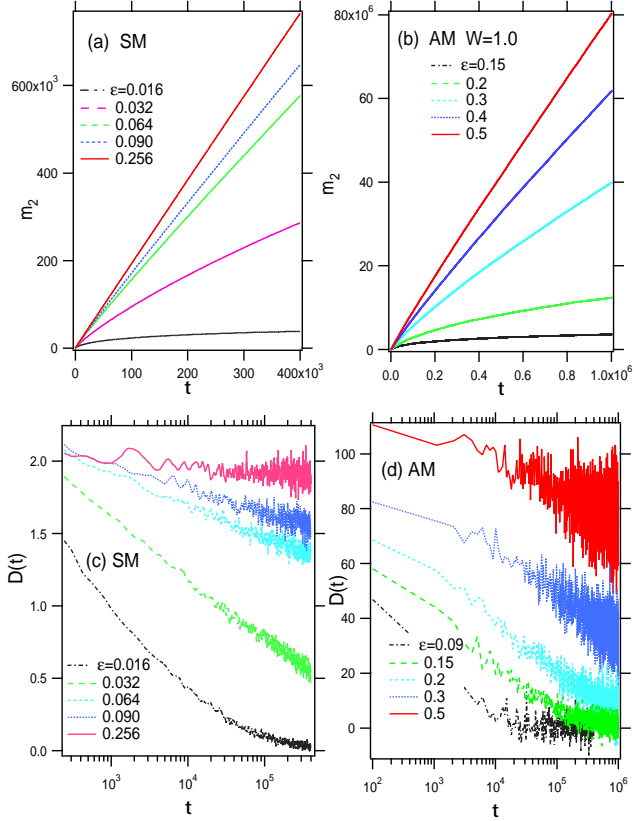}
\caption{(Color online)
The time-dependence of the MSD and the diffusion coefficient $D(t)$
for relatively large perturbation strength.
(a)Monochromatically perturbed SM for $K=3.1$ with some $\eps$s.
(b)Monochromatically perturbed AM for $W=1.0$ with some $\eps$s.
(c)(d) The $D(t)$ of the  SM  and the AM, respectively.
Note that the horizontal axes of the $D(t)$ are in logarithmic scale.
}
\label{fig:FigA}
\end{center}
\end{figure}

Figure \ref{fig:FigA}(a)  and (b) show the time-dependence of the MSD 
obtained for SM and AM by increasing the perturbation strength $\eps$, where 
$\hbar$ is fixed at a sufficiently small value for SM,  and $W$ is fixed to a 
large value such that $W \gg W^*$ for AM.
It becomes localized if $\eps$ is small, but as $\eps$ is increased larger, 
a diffusive behaviour emerges over a long time scale, which is 
apparently different from the monotonically localizing behaviour
typically seen in the Fig.\ref{fig:AMSM-fig1} in Sect.\ref{sec:model}.

To observe the diffusive behaviour qualitatively 
the time-dependent diffusion coefficient defined by
\beq
 D(t)=\frac{d\ovl{m}_2(t)}{dt},
\eeq
is convenient, where $\ovl{m}_2(t)$ indicates a smoothed curve over 
a sufficiently long section of time including $t$. 
Figures \ref{fig:FigA}(c) and (d) show 
the time-dependence of $D(t)$ for the SM and AM, respectively. 
The time-dependent diffusion coefficient decreases with time being accompanied
with fluctuation, but 
for sufficiently large $\eps$ the decrease of $D(t)$ is so slow that its variation
is detectable only in the logarithmic time scale. In the early stage it
seems to decreases linearly in logarithmic time scale. We remark that
in the case of SM, $D(t)$ agrees with the classical diffusion coefficient
$D_{cl}$ in the very initial stage, reflecting the quantum-classical 
correspondence within the Ehrenfest time.

\begin{figure}[htbp]
\begin{center}
\includegraphics[width=6.0cm]{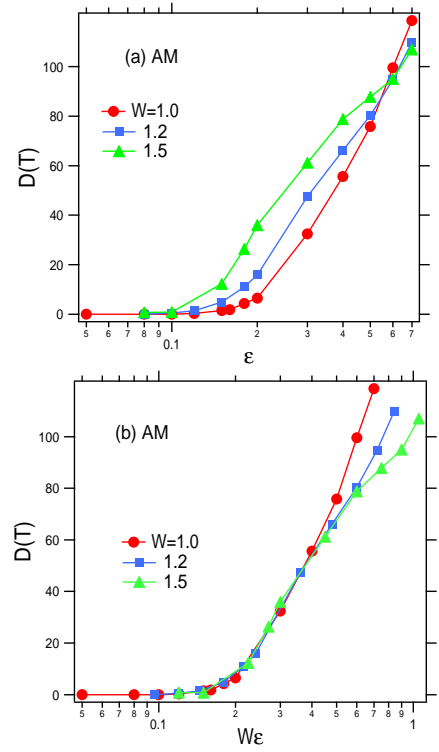}
\caption{(Color online)
(a)The time-dependent diffusion coefficient 
$D(T)$ at $T=1\times 10^6$ as a function of 
$\eps$ in the monochromatically perturbed AM with $W=1.0,1.2,1.5$.
(b) The $D(T)$ as a function of  $W\eps$.
Note that the horizontal axes are in logarithmic scale.
}
\label{fig:FigB-AM}
\end{center}
\end{figure}

\begin{figure}[htbp]
\begin{center}
\includegraphics[width=6.5cm]{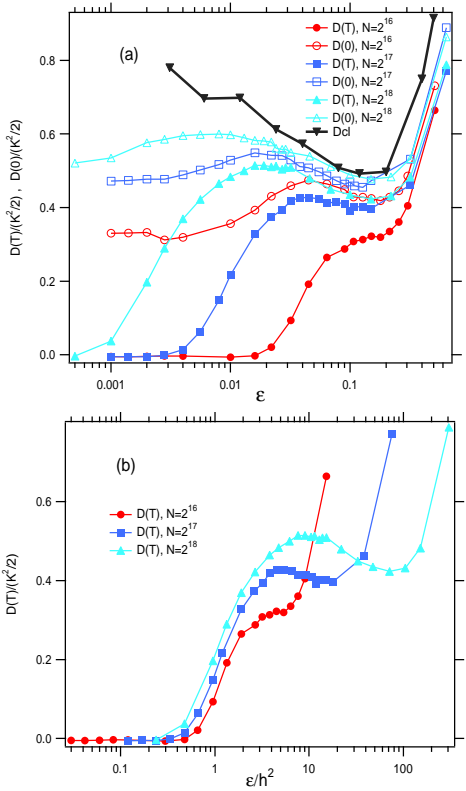}
\caption{(Color online)
(a)The time-dependent diffusion coefficient 
$D(T)/(K^2/2)$ at $T=4\times 10^5$ as a function of 
$\eps$ in the monochromatically perturbed SM 
with $\hbar=\frac{2\pi 1234}{2^{16}}$, 
$\hbar=\frac{2\pi 1234}{2^{17}}$, 
$\hbar=\frac{2\pi 1234}{2^{18}}$.  Here $K=3.1$ for all data.
(b) The $D(T)$ as a function of  $\eps/\hbar^2$.
Note that the horizontal axes are in logarithmic scale.
}
\label{fig:FigB-SM}
\end{center}
\end{figure}

Next we characterize the $\eps-$dependence of the time-dependent diffusion coefficient
$D (t)$. Consider the time evolution up to $t=T$. 
As is shown in Fig.\ref{fig:FigA} $D(t)$ takes the minimum value at $t=T$, and $D(T)=0$ 
means that the wavepacket have localized until $t=T$. 
We are concerned with $D(t)$ after a very long time evolution, but
we can not now specify the scale of $T$ on which the dynamics of 
localization process is characterized. At present,
we tentatively take the time scale $T$ as long as our numerical run 
time allows, and we represent the minimum value $D(T)$. $T$ is fixed at 
$5\times 10^5-10^6$ steps. The results are plotted as the function of
the perturbation strength $\eps$ for three values of $W$ (AM) and $\hbar$ (SM) 
in Figs.\ref{fig:FigB-AM}(a) and \ref{fig:FigB-SM}(a), 
respectively.
For the SM we also plot $D(0)$, which is the maximum value of $D(t)$ and mimics the classical diffusion 
constant $D_{cl}$, in order to show explicitly the range scanned by $D(t)$ 
in the time interval $0\leq t\leq T$. We plot also the classical diffusion coefficient
$D_{cl}$ as a function of $\eps$. 
It follows that both in SM and AM the $D(T)$ 
gradually rise with $\eps$, and later it increases rapidly.
  
In the AM, the $D(T)$ curve as a function of $\eps$ 
shifts upward with $W$, which is consistent with
the dependence of LL on $\eps$ and $W$ as discussed in the Sec.\ref{sec:AM}.
In fact, taking the parameter $W \eps$ instead of $\eps$
the curves of diffusion coefficient $D(T)$ in Fig.\ref{fig:FigB-AM}(a) 
are all well-overlapped, as shown in Fig.\ref{fig:FigB-AM}(b) if $\eps W$ is 
not too large.  
Recalling the result of the previous Sect.\ref{sec:AM} that the single parameter 
$W\eps$ controls the LL, it is quite
natural that the diffusion coefficients are also decided only by the combined
parameter $W\eps$.

Finally we discuss the very important feature of SM which is not seen in
AM. It is the existence of the semiclassical regime which emerges for small $\hbar$, 
as was shown in coupled SM.\cite{adachi88}.
As mentioned above, $D(0)$ mimics the classical diffusion coefficient, 
however, $D(t)$ decreases as is shown in Fig.\ref{fig:FigA}(c), and
the decaying rate in $\log t$ scale decreases with $\eps$, and there exists a 
characteristic value $\eps_c$ beyond which the decay becomes extremely 
small and so $D(T)\sim D(0)\sim D_{cl}$. Indeed, Fig.\ref{fig:FigB-SM}(a) 
exhibits that with increase in $\eps$, $D(T)$ increases rapidly, and beyond 
a certain $\eps=\eps_c$ it forms a plateau on which $D(T)$ keeps almost constant level. 
On the plateau the difference $D(T)-D(0)(\sim D_{cl})$ is small,  and
$D(T)$ approaches closer to $D_{cl}$ as $\hbar \to 0$. Thus we call the plateau 
as the ``classical plateau'' of the 
time-dependent diffusion coefficient. With further increase of $\eps$, 
the classical diffusion rate is enhanced and $D(T)$ 
takes off from the plateau following the enhanced $D_{cl}$ closely. 
Evidently, the classical plateau and the threshold $\eps_c$ 
shift toward smaller side of $\eps$ as $\hbar$ decreases. 
In the plots of  $D(T)$ as the function of the scaled parameter $\eps/\hbar^2$ shown in 
Fig.\ref{fig:FigB-SM}(b),  the left edges of the plateaus for different $\hbar$s coincide, 
which means that $\eps_c\propto \hbar^2$. Figure \ref{fig:FigB-SM}(b) also implies that
$D(T)$ is controlled by $\eps/\hbar^2$, as is the case of the localization
length of Eq.(\ref{eq:SM3}). Thus the quantum regime we introduced without definition
previously should be
\begin{eqnarray}
      \eps < \eps_c (= C \times \hbar^2), 
\end{eqnarray} 
where the constant $C$ depends on $K$ such as $C \propto K^2$, 
but we have not confirmed it yet.  
The localization characteristics of the SM discussed in the previous sections have been confirmed 
only in the quantum regime. It is still open to question whether or not the localization
characteristics represented by Eq.(\ref{eq:SM3}) is valid in the semiclassical regime.

The dynamical problems related to the localization process such as the existence of 
characteristic time 
leading to the localization and$/$or the existence of dynamical scaling property are 
still an interesting unclarified issue, particularly when the LL is extremely 
large for $\eps W \gg 1$ in AM with $W>W^{*}$ 
and for $\hbar\to 0$ in SM, respectively.

\section{Concluding remarks}
\label{sec:last}
We investigated the dynamical localization of 
the SM (standard map) and the AM (Anderson map) which are dynamically perturbed by a monochromatically
periodic oscillation, 
and the parameter dependence of the dynamical localization length has been clarified
by extensive numerical simulation and theoretical considerations. Under suitable 
conditions such systems could be identified with 
a 2DDS (two-dimensional disordered system) 
by using the so-called Maryland transform.

The dynamical localization length (LL)  was determined by the MSD computed by the numerical 
wavepacket propagation. We emphasize that the SM is treated in the quantum regime, 
where the perturbation strength $\eps$ is smaller than a characteristic value 
proportional to $\hbar^2$.
The LL increases exponentially with respect to $\eps$ in both perturbed SM and AM. 
It was further scaled by using the dynamical localization length of the unperturbed system 
in the case of SM, which is consistent with experimental results. On the other hand, in the 
case of AM, it was scaled by the disorder strength $W$.
There exists the threshold of the disorder strength $W^*$ at which a marked change of 
$W-$dependence of the dynamical localization length occurs. In the region, $W<W^*$,
the ordinary localization in 1DDS occurs, whereas new region, $W>W^*$, peculiar
to the quantum map emerged where the localization length increases with 
the disorder strength $W$ due to the kicked perturbation.

Next, we showed that all the numerically observed scaling characteristics mentioned
above can be reproduced in a {\it unified manner} 
by the self-consistent mean-field approximation 
theory developed for the localization of the 2DDS by introducing a new fundamental 
characteristic lengths as the cut-off length. 
This fact strongly suggests that the monochromatically perturbed SM and AM has essentially
the same physical origin for the exponentially enhanced localization length.

Finally the transient diffusive behaviour toward the dynamical localization
was investigated in the large limit of localization length. In both cases of 
SM and AM, the transient diffusion coefficient also follows the same scaling 
rule as the localization rule, but in the case of SM the ``classical plateau'' 
exists in the semiclassical regime, in which the compatibility of the quantum 
localization with the classical chaotic diffusion is quite interesting. 
Indeed, different type of localization which can not be captured by the
above mentioned ``unified picture'' may emerges in the semiclassical 
regime. 
These are interesting problems still open to question.

The Anderson map asymptotically approach to the original Anderson model 
in the limit of $\tau \to 0$ as mentioned in Sect.\ref{sec:model}.
Whether or not the result in 
this paper is true even in the time-continuous version (Anderson model) 
is also an interesting future problem.

\appendix

\section{other numerical data the perturbed AM}
\label{app:different-frequency}
\begin{figure}[htbp]
\begin{center}
\includegraphics[width=6.5cm]{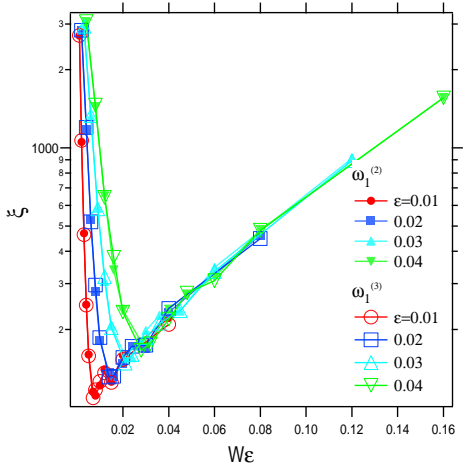}
\caption{(Color online)
The localization length as a function of 
scaled perturbation strength $\eps W$ in the monochromatically perturbed AM 
for $\eps=0.01,0.02,0.03,0.04$ 
with perturbation frequencies $\omega_1^{(2)}=\sqrt{2}-1$, $\omega_1^{(3)}=1+1/\sqrt{17}$.
All are displayed together.
}
\label{fig:AM-fig4}
\end{center}
\end{figure}

Figure \ref{fig:AM-fig4} displays the localization length as a function of 
scaled perturbation strength $\eps W$ in the monochromatically perturbed AM
 with the frequencies 
$\omega_1^{(2)}=\sqrt{2}-1$, $\omega_1^{(3)}=1+1/\sqrt{17}$
 different from one in the text.
At least the scaling of the localization length 
is a stable result even for the frequencies.
It follows that for $W>W^*$ the scaled $\eps-$dependence is overlapping
with each others.
Also, although not shown here, for $W<W^*$
the $W-$dependence of the localization length in the cases 
also behaves  similarly to that in the text.

\section{autonomous representation and Maryland transform}
\label{app:Maryland}
The eigenvalue problem Eq.(\ref{eq:eigen-value-problem})
 can be mapped into the tight-binding form 
by Maryland transform through the following states $|u_2\>$, $|u_1\>$ 
and Hermite matrices $\hatt$, $\hatw$;
\beq
|u_2\> &=& \e^{-i\hatB}\e^{-i\hatC}|u_1\>\\
|u_1\> &=& \e^{-i(\hatA-\gamma)}|u_2\>,
\eeq
\beq
\frac{1-i\hatt}{1+i\hatt} &=& \e^{-i\hatB}\e^{-i\hatC}, \\
\frac{1-i\hatw}{1+i\hatw} &=& \e^{-i(\hatA-\gamma)}.
\eeq
That is, 
\beq
\hatt(\hatq,\hatp,\phi) &=& -i\frac{1-\e^{-i\hatB}\e^{-i\hatC}}{1+\e^{-i\hatB}\e^{-i\hatC}},\\
\hatw(\hatp,\hatJ) &=& -i\frac{1-\e^{-i(\hatA-\gamma)}}{1+\e^{-i(\hatA-\gamma)}}= \tan \left[ \frac{(\hatA-\gamma)}{2} \right].
\eeq
Then the tight-binding form of the eigenvalue problem becomes 
\beq
 \left( \tan \left[ \frac{(\hatA-\gamma)}{2} \right]+ \hatt(\hatq,\phi) \right) |\ovl{u}\> &=& 0,
\label{eq:tan-1}
\eeq
where 
\beq
 (1+i\hatt)^{-1}|u_1\> &= & (1-i\hatt)^{-1}|u_2\>  \nn \\
                            &=& (1-i\hatt)^{-1}\e^{i(\hatA-\gamma)}|u_1\> \nn \\
                            & \equiv & |\ovl{u}\>,
\eeq
and
\beq
 |\ovl{u}\> = (|u_1\>+|u_2\>)/2.
\eeq
$\hatt =\tan [\hatC/2]$ when $\eps=0$.
We can select a convenient representation 
for the eigenvalue equation (\ref{eq:tan-1}).
In this case,  we dealt with the monochromatic perturbation 
in the autonomous representation, but  the extension to the case of multicolor perturbation 
can be easily done.

For the SM, the eigenvalue equation  in the representation by $u(m,j)=\<m,j|\ovl{u}\>$
based on the eigenstates $|m,j\>=|m\> \otimes |j\>$ of $\hatp$ and $\hatJ$, respectively,
is given by the following two-dimensional lattice system 
with aperiodic and singular on-site potential; 
\begin{widetext}
\beq
\tan \left[ \frac{\hbar^2m^2/2+j\omega_1\hbar}{2\hbar}\tau-\frac{\gamma}{2} \right] u(m,j)+ 
 \sum_{m',j'}\<m,j|\hatt|m',j'\>u(m',j')=0
\eeq
\end{widetext}
where the transfer matrix element is given 
by Eq.(\ref{eq:transfer-SM}) in the main text.

On the other hand, for the monochromatically perturbed AM, using 
\beq
\hatA &=& (Wv(\hatq)+\omega_1\hatJ )\tau/\hbar, \\
\hatB &=& \eps Wv(\hatq)(\cos \phi)\tau /\hbar, \\
\hatC &=& 2\cos (\hatp/\hbar)/\hbar 
\eeq
the eigenvalue equation (\ref{eq:tan-1}) is established as it is, then 
\begin{widetext}
\beq
 \left( \tan \left[ \frac{(\hatW(\hatq)+\omega_1\hatJ-\gamma)\tau}{2} \right]+ 
\hatt(\hatp,\hatq,\phi) \right) |\ovl{u} \> &=& 0.
\eeq
If we use the representation 
$u(n,j) (=(\<n|\otimes \<j|)|\ovl{u} \>)$
based on the eigenstates $|q=n\>$ of the site  $\hat{n}$ and $|j\>$ of the $\hat J$,
it becomes 
\beq
 \tan \left[ \frac{Wv_n+j\omega_1\hbar}{2\hbar}\tau-\frac{\gamma}{2}
 \right]u(n,j)+  
\sum_{n',j'}\< n,j|\hatt|n',j'\>u(n',j')=0, 
\eeq
where
\beq
\<n,j| \hatt|n',j'\> &=& \left<n,j \left|
i\frac{e^{-i\hatB}-e^{i\hatC}}
{e^{-i\hatB}+e^{i\hatC}}\right| n',j'\right> \\
&=& \left<n,j \left| i\frac{
e^{-i\eps Wv(\hatq)\cos\phi\tau/\hbar}-
e^{i2\cos(\hatp/\hbar)\tau/\hbar}
}
{e^{-i\eps Wv(\hatq)\cos\phi \tau/\hbar}+
e^{i2\cos(\hatp/\hbar)\tau/\hbar}
}\right| n',j'\right>.
\eeq
\end{widetext}
 This is the Maryland transformed eigenvalue equation including degrees of freedom 
of the monochromatic perturbation in the case of the AM.

\section*{Acknowledgments}
This work is partly supported by Japanese people's tax via JPSJ KAKENHI 15H03701,
and the authors would like to acknowledge them.
They are also very grateful to Dr. T.Tsuji and  Koike memorial
house for using the facilities during this study.



\begin{thebibliography}{00}
\bibitem{memo1}
The average over irrational random number is more efficient 
than the normal random number average,
because $U(s\tau,\varphi_0)$ is a periodic with respect to $\varphi_0$.
\end{thebibliography}
\end{document}